\documentclass[12pt]{iopart}

\begin{document}

\title{Effective medium theory of permeation through ideal polymer networks}
\author{Yong Wu}
\address{Department of Physics, Virginia Tech, Blacksburg, VA
24061-0435}
\ead{wuyong@vt.edu}

\begin{abstract}
The diffusion process through an ideal polymer network is studied by
applying the effective medium theory (EMT) to a lattice-gas model.
Polymers are modeled by random walks on the lattice bonds, across
which molecules can hop with a certain probability. The steady state
current of the system is calculated using the EMT and the results
are compared to the simulations.
\end{abstract}

\maketitle

\section{Introduction}
In this paper I apply the effective medium theory (EMT) to a
prototype lattice-gas model simulating the permeation process in
polymer networks. The permeation through polymer networks has been
studied in a number of experiments and simulations. Experiments are
usually performed on a thin polymer film subject to a constant gas
pressure. The current is measured at the steady state and its
relation to the temperature, the cooling rate, the size of molecules
and so on are studied.\cite{LaMaScZi03} A prototype lattice-gas
model, using random walks on lattice bonds to model the polymers,
has been constructed in simulation studies.\cite{ScGoZi05} This
simple model is equivalent to a resistor network, in which the
probability distribution of resistor on a bond is highly correlated
to its neighbors. The correlations differ this model from the
well-known randomly distributed resistor network and make it much
harder to study.

The EMT is a powerful tool to calculate the effective properties of
random media.\cite{Choy} It has been successfully used in
calculating effective conductance of the completely randomly
distributed resistor network when far away from percolation. In this
case, the Bruggeman formula of the EMT provides sufficiently
accurate results,\cite{Luck91} meaning that, when applied to our
polymer systems, the Bruggeman formula yields excellent
approximation when the polymer length is $1$. When the polymer
length gets longer, however, the error of the Bruggeman formula
grows due to increasing correlations, and the theory needs be
modified to account for the correlations. In this paper I develop a
method to calculate the effective diffusivity, which gives fairly
good results when the polymer length is relatively short (but
greater than $1$). Unfortunately, this method suffers from its fast
growing complexity when the correlation increases, and thus is
difficult to apply to very long polymer networks, or system close to
percolation.

\section{Model and Theory}
Consider a d-dimensional lattice. We use coordinates $\mathbf{x}=(x_1,...,x_d)$, where $%
x_1,\cdots ,x_d$ are integers, to label the lattice cells, and $\mathbf{e}_i$ (%
$1\leq i\leq d$) to denote the base vectors. Polymers are randomly
placed in the lattice, whose segments occupy the lattice bonds. In a
numerical simulation the polymers are generated by random walks. A
particle can occupy a lattice cell and can jump to one of its
neighboring cells, across the polymer segments sitting between the
two cells. The jumping probability depends on the number of polymer
segments. Similar to the Ohm's law in a resistor system, the steady
state current $J_i(\mathbf{x})$, which flows from
$\mathbf{x}+\mathbf{e}_i$ to $\mathbf{x}$, is related to the
concentration drop
$E_i(\mathbf{x})=c(\mathbf{x}+\mathbf{e}_i)-c(\mathbf{x})$ by
\begin{equation}
J_i(\mathbf{x})=\sigma _{ij}(\mathbf{x})E_j(\mathbf{x}),  \label{eq:ohm}
\end{equation}
where $\sigma _{ij}(\mathbf{x})$ is the diffusivity. The diffusivity
is a diagonal matric, whose diagonal elements $\sigma
_{ii}(\mathbf{x})$ are the jumping probability of particles across
the bond separating cell $\mathbf{x}+\mathbf{e}_i$ and cell
$\mathbf{x}$. Throughout this section we adopt the Einstein
summation convention that sums over repeated indices from $1$ to
$d$. The conservation of the number of molecules is written as
\begin{equation}
\nabla _i\left[ \sigma _{ij}(\mathbf{x})\Delta _jc(\mathbf{x})\right] =0,
\label{eq:laplace}
\end{equation}
where $\Delta _i$ is the forward finite difference such that, for any
function $f(\mathbf{x})$, $\Delta _if(\mathbf{x})=f(\mathbf{x}+\mathbf{e}%
_i)-f(\mathbf{x})$, and $\nabla _i$ is the backward finite difference
defined by $\nabla _if(\mathbf{x})=f(\mathbf{x})-f(\mathbf{x}-\mathbf{e}_i)$%
.

In a typical permeation experiment, a constant gradient of
concentration $\mathbf{E}_0$ is applied on the system.
It is convenient to define the effective diffusivity $\Sigma_{ij}$
by the proportionality between the average current and the applied
field.
\begin{equation}\label{eq:eff}
\langle J_i\rangle =\Sigma _{ij}\,(E_0)_j.
\end{equation}
where $\langle \cdot \rangle $ denotes the spatial average. Assuming the
system is isotropic, then we should have $\Sigma _{ij}=\sigma _{\mbox{eff}%
}\,\delta _{ij}$, where $\delta _{ij}$ is the Kronecker delta.

Let $\eta _{ij}(\mathbf{x})=\sigma _{ij}(\mathbf{x})-\Sigma _{ij}$. Eq.\ \ref
{eq:laplace} can be solved using the Green function technique.
\begin{equation}
c(\mathbf{x})=(E_0)_ix_i+\sum_{\mathbf{x}^{\prime }}G(\mathbf{\mathbf{x}%
},\mathbf{\mathbf{x}^{\prime }})\nabla _i^{\prime }\left( \eta _{ij}(%
\mathbf{\mathbf{x}^{\prime }})\Delta _j^{\prime
}{c(\mathbf{x}^{\prime })}\right) ,  \label{eq:cgreen}
\end{equation}
where $G(\mathbf{x},\mathbf{x}^{\prime })$ is the Green function satisfying
\begin{equation}
\sigma_{\mbox{eff}} \cdot \nabla _i\Delta _iG(\mathbf{x},\mathbf{x}^{\prime })=-\delta (%
\mathbf{x}-\mathbf{x}^{\prime })  \label{eq:green}
\end{equation}
and vanishing on the boundary.
Then $\mathbf{E}(\mathbf{x})$ is given by
\begin{equation}
E_i(\mathbf{x})=(E_0)_i+\sum_{\mathbf{x}^{\prime }}\,\tilde{G}_{ij}(\mathbf{x%
},\mathbf{x^{\prime }})\eta _{jk}(\mathbf{x^{\prime }})E_k(\mathbf{x^{\prime
}})  \label{eq:Egreen}
\end{equation}
where the $d\times d$ matrix $\tilde{G}(\mathbf{x},\mathbf{x^{\prime }})$ is
given by
\begin{equation}
\tilde{G}_{ij}(\mathbf{x},\mathbf{x^{\prime }})=-\Delta _i\Delta _j^{\prime
}G(\mathbf{x},\mathbf{x^{\prime }}).  \label{eq:tildeG}
\end{equation}

We also define susceptibility $\chi(\mathbf{x})$ as a matrix such that
\begin{equation}  \label{eq:sus}
\chi_{ij}(\mathbf{x})(E_0)_j = \eta_{ij}(\mathbf{x})E_j(\mathbf{x}),
\end{equation}
which describe the field $\eta(\mathbf{x})\mathbf{E}(\mathbf{x})$ as a
response to the applied field $\mathbf{E}_0$. Since $\langle\eta\mathbf{E}%
\rangle=\langle\mathbf{J}-\Sigma\mathbf{E}\rangle=\Sigma\left(\langle\mathbf{%
E}_0\rangle-\langle\mathbf{E}\rangle\right)=0$, the average of
susceptibility is zero,
\begin{equation}  \label{eq:self}
\langle\chi(\mathbf{x})\rangle = 0.
\end{equation}

Rewriting Eq.\ \ref{eq:Egreen} in terms of $\chi(\mathbf{x})$, we get the
following Dyson equation
\begin{equation}  \label{eq:solChi}
\chi(\mathbf{x})=\eta(\mathbf{x})+\eta(\mathbf{x})\sum_\mathbf{\mathbf{x}%
^\prime}\tilde{G}(\mathbf{x},\mathbf{x^\prime})\chi(\mathbf{x^\prime}),
\end{equation}
Decoposing the diagonal term
\begin{equation}  \label{eq:solChiDis}
\chi(\mathbf{x})=\eta(\mathbf{x})+\eta(\mathbf{x})\tilde{G}(\mathbf{x},%
\mathbf{x})\chi(\mathbf{x})+\eta(\mathbf{x})\sum_{\mathbf{x}\neq{\mathbf{x}%
^\prime}}\tilde{G}(\mathbf{x},\mathbf{x^\prime})\chi(\mathbf{x^\prime}).
\end{equation}
and introducing
\begin{equation}  \label{eq:EMT}
\chi^{E}(\mathbf{x}) = \lbrack1-\eta(\mathbf{x})\tilde{G}(\mathbf{x},\mathbf{%
x})\rbrack^{-1}\eta(\mathbf{x}),
\end{equation}
Eq.\ \ref{eq:solChiDis} can be simplified
\begin{equation}  \label{eq:solSim}
\chi(\mathbf{x}) = \chi^E(\mathbf{x}) + \chi^E(\mathbf{x})\sum_{\mathbf{x}%
\neq{\mathbf{x}^\prime}}\tilde{G}(\mathbf{x},\mathbf{x^\prime})\chi(\mathbf{%
x^\prime}).
\end{equation}
Expanding the right hand side of the above equation and taking the average,
we get
\begin{eqnarray}  \label{eq:opsol1}
\left\langle{\chi}(\mathbf{x})\right\rangle & = & \left\langle{\chi}^E(%
\mathbf{x})\right\rangle + \sum_{\mathbf{x}\neq\mathbf{x}^\prime}\left\langle%
{\chi}^E(\mathbf{x})\tilde{G}(\mathbf{x},\mathbf{x}^\prime){\chi}^E(\mathbf{x%
}^\prime)\right\rangle  \nonumber \\
& + & \sum_{\mathbf{x}\neq\mathbf{x}^\prime}\sum_{\mathbf{x}^\prime\neq%
\mathbf{x}^{\prime\prime}}\left\langle{\chi}^E(\mathbf{x})\tilde{G}(\mathbf{x%
},\mathbf{x}^\prime){\chi}^E(\mathbf{x}^\prime)\tilde{G}(\mathbf{x}^\prime,%
\mathbf{x}^{\prime\prime}){\chi}^E(\mathbf{x}^{\prime\prime})\right\rangle
+\cdots
\end{eqnarray}

We classify the cells by the number of polymer segments sitting on
its bonds. For example, when the dimension $d=2$, if each bond can
only be occupied by one polymer segment at most, and if the
probability of jumping across a polymer segment is $q$, then there
are $4$ different types of cells
whose diffusivity matrices are
\begin{equation}
\sigma _1=\left(
\begin{array}{cc}
1 & 0 \\
0 & 1
\end{array}
\right) \sigma _2=\left(
\begin{array}{cc}
q & 0 \\
0 & 1
\end{array}
\right) \sigma _3=\left(
\begin{array}{cc}
1 & 0 \\
0 & q
\end{array}
\right) \sigma _4=\left(
\begin{array}{cc}
q & 0 \\
0 & q
\end{array}
\right) ,  \label{eq:4cell}
\end{equation}
respectively. Noticing that $\chi ^E(\mathbf{x})$ is determined by the type
of cell at $\mathbf{x}$, and in a homogeneous system, $\tilde{G}(\mathbf{x},%
\mathbf{x}^{\prime })$ only depends on $\mathbf{x}-\mathbf{x}^{\prime }$,
Eq.\ \ref{eq:opsol1} can thus be rewritten in terms of statistics of
different type of cells
\begin{eqnarray}
\langle {\chi }\rangle  &=&\sum_if_0(i)\,{\chi }_i^E+\sum_{i,j}\sum_{\mathbf{%
x^{\prime }}\neq \mathbf{x}}f_1(\mathbf{x}^{\prime }-\mathbf{x};i,j)\,{\chi }%
_i^E\,\tilde{G}(\mathbf{x}^{\prime }-\mathbf{x})\,{\chi }_j^E  \nonumber
\label{eq:stat} \\
&+&\sum_{i,j,k}\sum_{\mathbf{x^{\prime }}\neq \mathbf{x}}\sum_{\mathbf{x}%
^{\prime }\neq \mathbf{x}^{\prime \prime }}f_2(\mathbf{x}^{\prime }-\mathbf{x%
},\mathbf{x}^{\prime \prime }-\mathbf{x}^{\prime };i,j,k)\,{\chi }_i^E\,%
\tilde{G}(\mathbf{x}^{\prime }-\mathbf{x}){\chi }_j^E\,\tilde{G}(\mathbf{x}%
^{\prime \prime }-\mathbf{x}^{\prime })\,{\chi }_k^E  \nonumber \\
&+&\cdots .
\end{eqnarray}
where $f_0(i)$ is the fraction of the type $i$ cells, $f_1(\mathbf{x}%
^{\prime }-\mathbf{x};i,j)$ is the joint probability distribution of seeing
a type $i$ cell and a type $j$ cell resting at $\mathbf{x}$ and $\mathbf{x}%
^{\prime }$, respectively, and $f_2$, $f_3$, $\cdots $ are defined
similarly. The function $\tilde{G}(\mathbf{x})=\tilde{G}(0,\mathbf{x})$ can
be easily calculated using the continuous Green function $G_c(\mathbf{x},%
\mathbf{x}^{\prime })$ of the entire $d$-dimensional space
\begin{eqnarray}
\tilde{G}_{ij}(\mathbf{x}) &=&\Delta _i\nabla _jG(0,\mathbf{x})
\label{eq:dgreed2D} \\
&=&\Delta _i\left[ \int_{x_1-\frac 12}^{x_1+\frac 12}\,dx_1^{\prime }\cdots
\int_{x_d-\frac 12}^{x_d+\frac 12}\,dx_d^{\prime }\,\frac{\partial G_c(0,%
\mathbf{x}^{\prime })}{\partial x_j^{\prime }}\delta \left( x_j^{\prime
}-\left( x_j-\frac 12\right) \right) \right] ,  \nonumber
\end{eqnarray}
where $\delta (x)$ is the delta function.

In principle, the probability distributions $f_i$ can be determined by doing
statistics for a numerically generated polymer system. The computer resource
required to do the statistics, however, increase exponentially with $i$ and $%
|\mathbf{x}-\mathbf{x}^\prime|$, $|\mathbf{x}^\prime-\mathbf{x}%
^{\prime\prime}|$, etc. Therefore in practice we are only able to calculate
the right hand side of Eq.\ \ref{eq:stat} up to $3$ terms, with the
constrictions that $|\mathbf{x}-\mathbf{x}^\prime|<R$, $|\mathbf{x}^\prime-%
\mathbf{x}^{\prime\prime}|<R$, $\cdots$, where $R\geq1$ is some constant
integer we pick. Neglecting of the correlations outside the range $R$ and
higher order terms are justified by the fact that $\tilde{G}(\mathbf{x})$
decreases as $|\mathbf{x}|^{-d}$.

Combining Eq.\ \ref{eq:stat} and Eq.\ \ref{eq:self} we get an equation that
can be used to solve for the effective diffusivity $\Sigma _{ij}$. If only
the first order term of the right hand side of Eq.\ \ref{eq:stat} is kept,
then it is only needed to solve $\langle \chi ^E\rangle =0$, which yields
the classical EMT solution for the effective diffusivity. Noticing that $%
\tilde{G}_{ij}(0)=-(1/d)\delta _{ij}$, the equation produces the well-known
Bruggeman formula
\begin{equation}
\frac{(1-p)(1-\sigma _{\mbox{eff}})}{1+(d-1)\sigma _{\mbox{eff}}}+\frac{%
p(q-\sigma _{\mbox{eff}})}{q+(d-1)\sigma _{\mbox{eff}}}=0,
\label{eq:classical}
\end{equation}where $p$ is the fraction of occupied bonds in the
system.

In a previous study it has been shown that for a system with the
polymer length $\ell =1$, Eq.\ \ref{eq:classical} gives an excellent
approximation to $\sigma _{\mbox{eff}}$.\cite{Luck91} However, if
the polymer length $\ell
>1$, the higher order terms, which describe the correlation between cells,
significantly revise the results. It is also worthy of noting that this
method of calculating the effective diffusivity works best when the system
is far away from the percolation threshold, which means either the
temperature has to be high, or the mass density be very low. Only in this
regime the correlation is not crucial. and this method can be a good
approximation.

For long polymers or system not so far away from percolation, or, in
other word, when correlations are important, one could improve the
result of the EMT by including more higher order terms in Eq.\
\ref{eq:stat}. This is illustrated by our calculation result shown
in Table\ \ref{tab:acu25}, where we list the current at $q=0.1$ and
$p=0.25$ calculated through Eq.\ \ref {eq:stat}, using up to $3$
higher order terms in the expansion, in comparison with both the
current measured by numerical simulations (for details of the
simulation, see \cite{WSZ07}) and the current calculated by the
classical EMT approximation. The correlation range $R$ used in the
calculation is $3$. Only single occupation on bonds is allow in the
calculation, because multiple occupation produces too many types of
cells to do statistics for $R>1$ or include more than one higher
order terms.

\begin{table}[tbp]
\caption{Current measured in simulations $J_{\mbox{sim}}$ and
current $J_{i}$ calculated by Eq.\ \ref{eq:stat}, which includes up
to $i$ higher order terms, as a function of the polymer length $l$.
Also shown in the brackets are the errors of each calculated current
relative to the measurement. Note that $J_0$ gives the classical EMT
solution, which is independent of the polymer length. The
probability to jump across an occupied bond is $q=0.1$, and the
occupation is $p=0.25$.} \label{tab:acu25}\centering
\begin{tabular*}{0.9\textwidth}{@{\extracolsep{\fill}}c|lllll}
\hline\hline
$l$ & $J_{\mbox{sim}}$ & $J_0$ & $J_1$ & $J_2$ & $J_3$ \\
\hline
1 & 156.4 & 155.7 (-0.45\%) & 155.7 (-0.45\%) & 155.9 (-0.32\%) & 155.9
(-0.32\%) \\
2 & 149.6 & 155.7 (4.1\%) & 151.5 (1.3\%) & 151.3 (1.1\%) & 151.1 (1.0\%) \\
4 & 144.5 & 155.7 (7.8\%) & 150.5 (4.2\%) & 148.9 (3.0\%) & 148.7 (2.9\%) \\
8 & 141.0 & 155.7 (10.4\%) & 149.7 (6.2\%) & 147.1 (4.3\%) & 146.7 (4.0\%)
\\
16 & 138.9 & 155.7 (12.1\%) & 149.4 (7.6\%) & 145.8 (5.0\%) & 145.4 (4.7\%)
\\
32 & 137.7 & 155.7 (13.1\%) & 150.1 (9.0\%) & 146.0 (6.0\%) & 145.5 (5.7\%)
\\
64 & 137.2 & 155.7 (13.5\%) & 150.6 (9.8\%) & 146.1 (6.5\%) & 145.6 (6.1\%)
\\ \hline\hline
\end{tabular*}
\end{table}

The results clearly show that the classical EMT yields extremely
accurate result for $\ell =1$, but produces considerable errors when
$\ell >1$. Taking account of correlations by including the higher
order terms can significantly improve the calculation.

\section{Summary and discussion}

\label{sec:sum} In this paper I have used the EMT to calculate the
effective diffusivity of a lattice-gas model simulating permeation
through polymer networks. Such a model is equivalent to resistor
network, which is described by a Laplace Equation at the steady
state. The solution of this equation can be expressed by a Dyson
expansion. One needs to include higher order terms in the expansion
to account for the correlations of the distribution of polymer
segments. When system is far away from percolation, we have
calculate the steady state current using up to $4$ terms in the
expansion and get reasonably good result compared to simulations.
The error of the calculation increases with the polymer length,
since longer polymers have stronger correlations.

From Table\ \ref{tab:acu25} one can see that the current is
decreasing function of the polymer length. It can be qualitatively
explained by considering the blob (the set of cells that carry
current at $q=0$) size, which also decrease with the polymer
length.\cite{WSZ07} A possible future work would be using the EMT to
provide some quantitative insight to this phenomenon.

\section{Acknowledgment}

This research is supported in part by the US National Science
Foundation through grant DMR-0414122.

\end{document}